# Sex chromosome stability and turnover across vertebrates: a developmental gene regulatory network perspective

Wen-Juan Ma[1†], Ricard Fontserè[1], Tristan Cornelis[1], Paris Veltsos[1], Qi Zhou[2,3,4]

1. Research Group of Ecology, Evolution and Genetics, Biology Department, Vrije Universiteit Brussel, Brussels, Belgium
2. Center for Reproductive Medicine, Second Affiliated Hospital of Zhejiang University School of Medicine, Life Sciences Institute, Zhejiang University, Hangzhou, China
3. Zhejiang Key Laboratory of Molecular Cancer Biology, Life Sciences Institute, Zhejiang University, Hangzhou, China
4. State Key Laboratory of Transvascular Implantation Devices, Second Affiliated Hospital of Zhejiang University School of Medicine, Life Sciences Institute, Zhejiang University, Hangzhou, China

[†]Correspondence:

wen-juan.ma@vub.be

# Abstract

Sex chromosomes have evolved repeatedly across vertebrates, yet their evolutionary fates differ strikingly among lineages. In mammals and birds, ancient XY and ZW systems have remained stable for over 100 Myr, whereas many fishes, amphibians, and non-avian reptiles possess largely homomorphic sex chromosomes that frequently turn over. Classical explanations for this contrast emphasize sex chromosome degeneration or population genetic processes and include the evolutionary trap hypothesis, sexually antagonistic selection, mutation load accumulation, genetic drift, selfish genetic elements and chromosome-specific drift. These models do not fully account for the difference in turnover rate between vertebrate clades. Here we propose the developmental gene regulatory network (GRN) lock-in hypothesis as a complementary framework. We compile vertebrate case studies of sex chromosome turnover and synthesise comparative developmental evidence on sex determination and dosage regulation. In therian mammals and birds, sex is specified by an early, highly penetrant, genetic signal acting in somatic cells, such as *Sry-Sox9* regulation or *Dmrt1* dosage, operating within a narrow and thermally buffered development window. This signal is embedded in canalised GRNs and coupled to chromosome-scale or otherwise strong dosage compensation, potentially making invasion by alternative sex-determining loci developmentally costly. By contrast, many ectothermic vertebrates possess more flexible and integrative sex-determining GRNs, in which genetic, epigenetic, germ cells and environmental inputs can act over broader sensitive time periods, often with local or gene-specific dosage regulation and regulatory plasticity near key sex-determining pathway nodes. Such architectures likely provide more entry points for new master sex-determining loci. We outline testable predictions of this framework and argue that integrating developmental biology, molecular mechanisms, comparative genomics, and population genetics can clarify when sex chromosomes become evolutionarily stable and when they remain prone to repeated turnovers.

## Keywords:

Sex chromosome turnover, sex chromosome degeneration, vertebrates, developmental gene regulatory network lock-in, Sertoli cell, dosage compensation

## Significance statement

Why sex chromosomes remain stable for more than 100 million years in some vertebrate lineages, yet turn over repeatedly in others, remains a major unresolved question in evolutionary biology. This review argues that the answer depends not only on population genetic processes, but also on the developmental architecture of sex determination and gene dosage compensation. By synthesizing comparative evidence across vertebrates, we introduce the developmental gene regulatory network (GRN) lock-in hypothesis, which proposes that strongly canalized sex-determining systems favour long-term sex chromosome stability, whereas more flexible and integrative systems may permit repeated turnovers. This framework links sex chromosome evolution with developmental biology and generates testable predictions for when sex chromosome systems are evolutionarily stable or remain labile.

## 1. Sex chromosome turnover across vertebrates: concepts, patterns and proposed mechanisms

Sex chromosomes, which usually carry the master sex-determining gene (MSD), have independently evolved many times across the Tree of Life (Bachtrog et al. 2014; Peichel 2017; Wilson Sayres 2018; Vicoso 2019; Ma and Rovatsos 2022; Veltsos et al. 2024). They follow strikingly different evolutionary trajectories in different vertebrate lineages. In mammals and birds, sex chromosomes are deeply conserved and sex-limited chromosomes (i.e. Y or W) are highly degenerated with many functional genes having become lost, whereas in most amphibians, many teleost fishes and non-avian reptiles they are largely homomorphic. One explanation for homomorphy is frequent sex chromosome turnover, i.e. replacement or reconfiguration of sex chromosomes among closely related taxa (Gamble et al. 2015; Lubieniecki et al. 2015; Jeffries et al. 2018; Palmer et al. 2019; Saunders 2019; Vicoso 2019; Furman et al. 2020; El Taher et al. 2021; Ma and Veltsos 2021). Sex chromosome turnover can be described along two independent axes (Figure 1a) (Saunders 2019; Veltsos et al. 2024). First, the new sex-determining region may arise on the same linkage group as the ancestral sex chromosome (homologous turnover) or on a different linkage group (heterologous turnover). Second, the turnover may either preserve the heterogametic sex (cis-heterogamety, e.g. XY-to-XY or ZW-to-ZW) or change it (trans-heterogamety, e.g. XY-to-ZW or ZW-to-XY). These two axes generate four possible combinations: homologous cis-heterogamety, heterologous cis-heterogamety, homologous trans-heterogamety, and heterologous trans-heterogamety (Figure 1a). All four combinations have been reported in nature, including homologous trans-heterogamety in Japanese frog *Glandirana rugosa* (Katsura et al. 2026).

Several evolutionary forces have been proposed to explain why new sex-determining systems spread or why established sex chromosomes resist replacement and turnover (Saunders 2019; Vicoso 2019). i) Sexually antagonistic selection model: a newly evolved sex-determining allele may spread if it becomes linked to alleles that are beneficial to one sex but deleterious to the other, thereby increasing sex-specific fitness and potentially promoting turnover (Van Doorn and Kirkpatrick 2007; Van Doorn and Kirkpatrick 2010). This model could drive turnovers mediated by novel masculinising or feminising mutations and predicts roughly equal rates between XY and ZW transitions. ii) Deleterious mutation load model: As recombination is suppressed around the sex-determining region, deleterious mutations would accumulate on the non-recombining Y/W chromosomes. Replacing the old sex chromosomes while retaining the same heterogametic sex allows for loss of the mutation-loaded Y or W, favouring cis-heterogametic transitions such as XY-to-XY or ZW-to-ZW turnover (Blaser et al. 2013). iii) Genetic drift model: new sex-determining alleles can drift to fixation, especially in small populations. Theory predicts that turnovers that retain the same heterogametic sex (cis transitions) should occur 2–4 times more often than switches between XY and ZW systems, partly because X or Z chromosomes are initially more frequent than the Y or W chromosomes (Veller et al. 2017; Saunders et al. 2018). iv) Sex-ratio selection model: meiotic drive, biased inheritance, endosymbionts, hybridization, or linkage between a new sex-determining allele and a meiotic driver can skew sex ratios. Selection may then favour alternative sex-determining systems that restore balanced sex ratios, potentially promoting transitions between XY and ZW systems (Werren and Beukeboom 1998; Kozielska et al. 2010; Ubeda et al. 2015; Franchini et al. 2018; Dixon et al. 2019; Katsumi et al. 2022; Miura, Shams, et al. 2024). v) Chromosome-specific drift under stabilizing selection: stabilizing selection on polygenic traits can allow chromosomes to accumulate chromosome-specific contributions to complex traits. Novel sex-determining genotypes generated during turnover may thus have reduced fitness, creating a general barrier to turnover and favouring long-term stability of established sex chromosome systems (Muralidhar 2026). These forces are not mutually exclusive. For example, the hot-potato model combines deleterious mutation accumulation with sexually antagonistic selection to generate recurrent cycles of sex chromosome turnover (Blaser et al. 2014).

Sex chromosome turnover can occur through both *de novo* changes within a lineage and the introgression of sex-determining regions between lineages. At the proximate genetic level, two mechanisms have been proposed for *de novo* turnover. First, the MSD or region can be translocated to a different chromosome both within and across species, for example via chromosomal rearrangements or transposable elements. The mechanism has been identified in several cases such as in the housefly (Kozielska et al. 2008; Meisel et al. 2016), salmonid fish (Phillips 2013), strawberry (Tennessen et al. 2018), Salicaceae plants (Yang et al. 2021), takifugu pufferfish (Kabir et al. 2022) and *Silurus* catfish (Zheng et al. 2025). Second, mutations at a downstream gene of the sex-determining pathway can move up and take over as the MSD (Jeffries et al. 2018; Adolfi et al. 2021; Wang et al. 2026). This mechanism has been directly supported in European toads (Dufresnes et al. 2025). In addition to these *de novo* origins, hybridization and introgression can promote sex chromosome turnover by transferring already-established sex-determining locus or sex-linked regions between lineages. Such introgressed region may either replace the current sex-determining system directly or, following chromosomal rearrangement, contribute to the formation of a novel sex chromosome (Franchini et al. 2018; Dixon et al. 2019; Katsumi et al. 2022; Miura, Shams, et al. 2024). Thus, introgression does not necessarily represent a distinct molecular origin of a sex-determining gene, but it provides an important evolutionary route by which sex chromosome turnover can occur across populations or species.

Sex chromosome turnover has been documented in many teleost fishes, amphibians, and non-avian reptiles, but it appears very rare in mammals and birds (Figure 1b) (Palmer et al. 2019; Saunders 2019; Vicoso 2019; Ma and Veltsos 2021; Veltsos et al. 2024; Wang et al. 2024). Comparative analyses across vertebrates suggest that transitions between XY and ZW systems occur at broadly similar rates in amphibians and squamates, whereas fishes show particularly frequent sex chromosome turnover, with transitions toward XY systems inferred to be especially common (Pennell et al. 2018). Complementary analyses in fishes suggest an asymmetry in sex-chromosome differentiation, with ZW systems tending to be more differentiated than XY systems (Sember et al. 2021). Wang et al. (2024), using an updated dataset, recovered the same broad pattern, with XY systems significantly more often homomorphic and ZW systems more often heteromorphic. Together, these patterns suggest that differences in the pace of sex-chromosome differentiation may influence the likelihood of turnover, with relatively undifferentiated XY systems potentially experiencing higher turnover rates. In our non-exhaustive compilation of case studies on sex chromosome turnover across vertebrates (Figure 1b), we focused on cases in which the sex-determining system was inferred to have changed, either through a change in heterogamety or replacement of the sex-determining region/linkage group. For consistency, we distinguish structural expansion of an existing sex-chromosome system through sex chromosome–autosome fusion from turnover of the primary sex-determining system. We therefore classify a fusion as turnover only when there is evidence for a change in heterogamety, replacement or relocation of the primary sex-determining locus/region, or a demonstrated change in the chromosome/linkage group carrying the primary sex-determining system, including where chromosome homology is resolved cytogenetically but the underlying MSD remains unknown. Fusions that simply recruit autosomal material to an otherwise retained sex-determining system are treated here as neo-sex chromosome formation rather than turnover (Supplementary Table S1). Based on these criteria, 65.6% (42/64) of reported cases concern fishes, 18.7% (12/64) amphibians and 10.9% (7/64) non-avian reptiles. In contrast, turnover of the primary sex-determining system appears to be exceptionally rare in birds and mammals. However, multiple rodent lineages have lost the ancestral Y chromosome and *Sry*-dependent sex determination, including *Ellobius* mole voles and *Tokudaia* spiny rats, which have evolved XX/XX or XO/XO systems with alternative, *Sry*-independent mechanisms of sex determination (Parma et al. 2016; Schartl and Herpin 2018) (Supplementary Table S1). Reported sex chromosome turnover events span at least 23 families in fishes, 27 in non-avian reptiles, 8 in amphibians, 2 in mammals and 0 in birds. The highest numbers of reported turnover events occur in Cichlidae among teleost fishes (30 out of 244 studied species), Ranidae among amphibians (>13 of 28 species) and Gekkota among non-avian reptiles (at least 12 of 63 studied species) (Figure 1b). Although part of the diversity in Gekkota likely reflects independent origins of GSD from ancestral ESD rather than turnover (Augstenová et al. 2021), at least 12 transitions among established GSD systems remain after excluding such cases. This represents a conservative minimum because sex-chromosome homology remains unresolved in many gekkotan lineages. Previous studies have identified cichlid fishes and ranid frogs as vertebrate lineages with exceptionally rapid sex chromosome turnover. In Lake Tanganyika cichlids, turnover rates were estimated at ~0.259 turnovers per Myr of divergence, while at least 13 turnover events were inferred among 28 ranid frog species within ~50 Myr of divergence (Jeffries et al. 2018; El Taher et al. 2021).

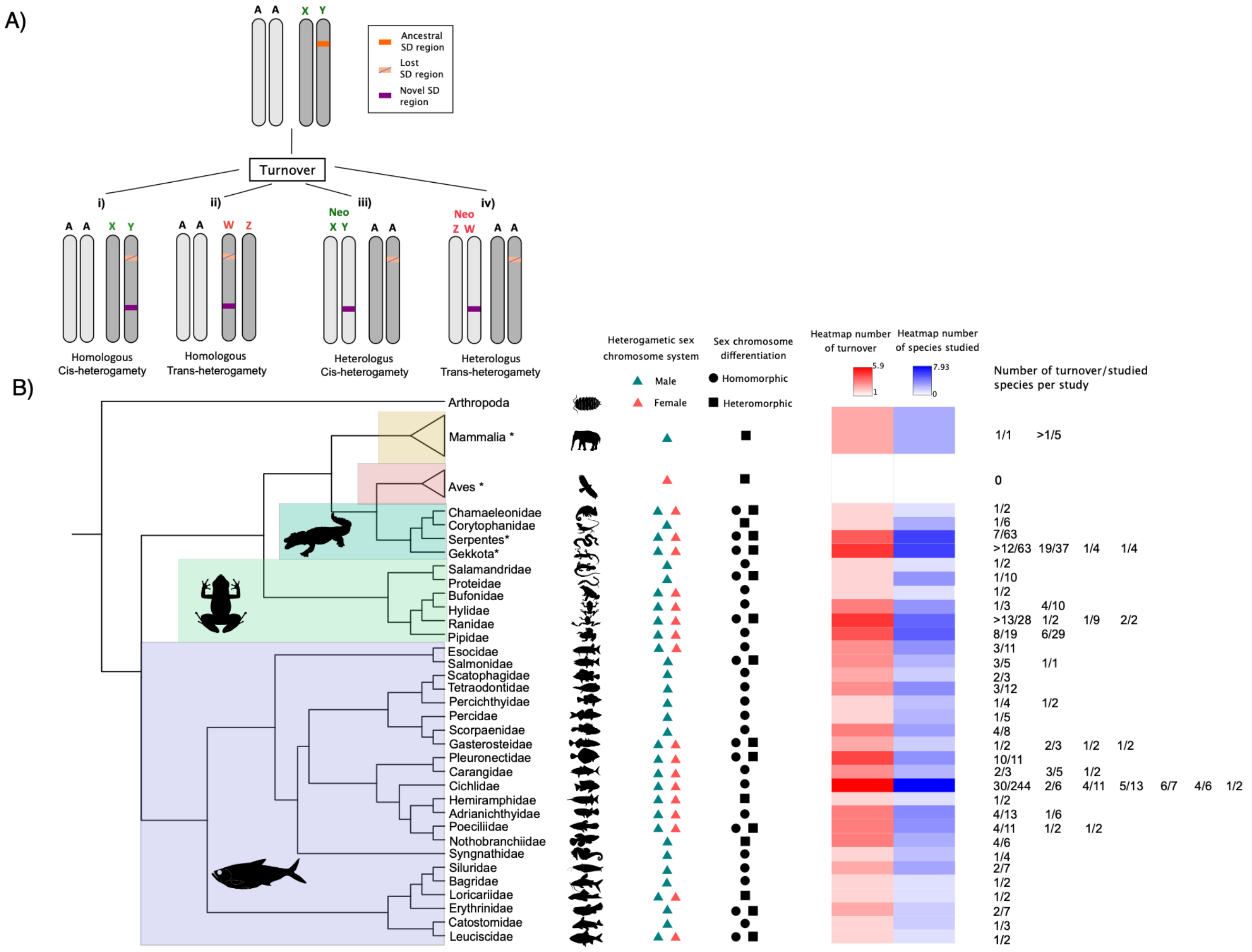

**Figure 1**. **Types and phylogenetic distribution of reported sex chromosome turnover across vertebrates.** (A) Schematic of four turnover types classified by two axes: whether the new sex-determining region arises on the same linkage group (homologous) or a different linkage group (heterologous), and whether the heterogametic sex is preserved (cis-heterogamety) or changed (trans-heterogamety). Orange indicates the ancestral sex-determining region, hatched orange its loss or replacement, and purple the novel sex-determining region. (B) Reported turnover events across vertebrate families, summarized from the literature in Supplementary Table S1. The phylogenetic tree was generated using http://timetree.org, and animal silhouettes were obtained from open-access PhyloPic resources: https://www.phylopic.org. Heatmaps show log-transformed numbers of turnover events and species studied per family; raw counts are shown as turnover events/studied species. Heterogametic sex and sex chromosome differentiation, homomorphic or heteromorphic, were scored from the original studies. XY and ZW denote male and female heterogamety, respectively, and do not necessarily imply simple two-

chromosome systems. Asterisks indicate cases scored at higher taxonomic levels where family-level resolution was unavailable.

A long-standing class of explanations behind the long-term stability of sex chromosome systems is the evolutionary trap hypothesis, which emphasises constraints caused by degeneration of the sex-limited chromosomes (Y/W). Once recombination is suppressed around the master sex-determining locus, Y/W chromosomes accumulate repetitive DNA, specialised regulatory features and even gene loss. Consequently, evolving a new master sex-determining locus on a different chromosome is expected to generate maladaptive intermediates (e.g. unbalanced gene dosage, disrupted meiotic pairing, reduced fertility) and may expose the deleterious effects of YY or WW genotypes (Pokorná and Kratochvíl 2009; Charlesworth 2021a). Degeneration alone cannot account for the observed diversity of sex chromosome dynamics. Some lineages with relatively undifferentiated sex chromosomes show little or no evidence of turnover (Zhou et al. 2014; Iwasaki et al. 2021; Kostmann et al. 2021; Kuhl et al. 2021; Han et al. 2022; Coffing et al. 2025; Niwa et al. 2025), whereas others, including several teleost and reptile clades, show repeated changes in sex chromosomes despite clear Y/W degeneration (Darolti et al. 2019; Nielsen et al. 2019; Rovatsos et al. 2019; Vicoso 2019). Moreover, formal comparative analyses across vertebrates do not strongly support a simple irreversible evolutionary trap: estimated turnover rates were found to be similar for homomorphic and heteromorphic systems in both fishes and amphibians (Pennell et al. 2018), suggesting that substantial sex chromosome differentiation does not necessarily preclude turnover. In fishes, a complementary pattern has emerged: female-heterogametic (ZW) systems tend to be more differentiated than male-heterogametic (XY) systems, with XY systems more often remaining homomorphic (Sember et al. 2021; Wang et al. 2024). This asymmetry may provide greater opportunity for turnover in relatively undifferentiated XY systems. In addition, cases of sex chromosome dedifferentiation show that previously differentiated sex chromosomes can, under some circumstances, return toward homomorphy (Kratochvíl, Stöck, et al. 2021).

These contrasts indicate that degeneration and the classical evolutionary forces above are only part of a broader picture. Comparative developmental and transcriptomic data point to an additional dimension: the architecture and dynamics of the developmental gene regulatory network (GRN). In mammals and birds, alternative splicing of certain factors (e.g. *Wt1*, *Fgfr2*, *Lef1*) is extensive and functionally important, but current evidence places these splice decisions in non-essential node and plays little role in affecting transcriptionally defined master switch initiated within the germline somatic cells (*Sry→Sox9* or *Dmrt1* dosage), which operates in a narrow development window in the thermally buffered endothermic embryos (Hammes et al. 2001; Kim et al. 2007; Zhao et al. 2018; Gómez-Redondo et al. 2021; Ioannidis et al. 2021). In ectothermic vertebrates with labile sex determination, available evidence suggests that alternative splicing can sometimes act close to the sex-determination decision point and be modulated by temperature or other external cues, although direct evidence remains taxonomically limited. In reptiles with temperature-dependent sex determination, *Rbm20* controls *Wt1* isoform usage during male pathway development, and temperature-sensitive splicing of *Kdm6b* or *Jarid2* has been implicated in sex reversal. In sex-changing fishes, environmentally responsive isoforms of genes such as *Dmrt1* and *Cyp19a1a* may also contribute to sexual fate regulation (Domingos et al. 2018; Ge et al. 2018; Whiteley et al. 2022; Sun et al. 2025). More broadly, alternative splicing of key sex-determining genes, including *Dmrt1*, has been reported across mammals, birds, reptiles, and fishes, whereas comparable evidence remains scarce in amphibians (Augstenová and Ma, 2025). Recent long-read transcriptomic analyses in guppy further revealed extensive sex-biased isoform usage, particularly in gonadal tissue, although much of the elevated male isoform diversity may reflect transcriptional noise rather than functional isoform diversification (Sherin et al. 2026). Beyond sex determination itself, sex chromosome differentiation may also affect splicing in other evolutionary contexts; for example, defective splicing associated with Y chromosome differentiation has been implicated in hybrid incompatibility in *Drosophila*, and was found in non-recombining algal mating-type chromosomes (Condon et al. 2026; Fontan et al. 2026).

Several previous reviews and theoretical studies have addressed why sex chromosomes turn over rapidly in some lineages but remain stable in others. Classical and comparative syntheses have emphasized recombination suppression, sexually antagonistic selection, mutation load, degeneration, and dosage compensation, while also showing that many fishes, amphibians and reptiles deviate from this canonical trajectory (Bachtrog et al. 2014; Beukeboom and Perrin 2014; Vicoso 2019; Furman et al. 2020; Ma and Veltsos 2021). More recent work further highlights that homomorphic sex chromosomes can persist for long periods, degeneration is not inevitable, and turnover is frequent in fishes, amphibians, and some reptiles but rare in mammals and birds (Furman et al. 2020; Kratochvíl et al. 2021; Ma and Veltsos 2021; Perrin 2021; Ma and Rovatsos 2022). Population-genetic models have identified conditions under which new sex-determining loci can invade, but developmental mechanisms remain insufficiently integrated (van Doorn 2014). Recent frameworks have begun to bridge this gap by proposing that sex-determination network architecture, epistasis, and chromosome-specific drift under stabilizing selection can constrain or bias sex chromosome replacement (Tao et al. 2021; Muralidhar 2026).

Building on these ideas, we propose the developmental gene-regulatory network (GRN) lock-in hypothesis as a framework for understanding why sex chromosomes are stable in some vertebrate lineages but labile in others (Figure 2). We suggest that turnover becomes less likely when sex determination is controlled by an early-acting, highly penetrant master regulator and when this regulator becomes integrated with dosage compensation, sex-linked gene regulation, and transcript-level regulatory mechanisms. This may characterize long-term stable sex chromosome systems such as therian mammals, birds, and sturgeons (Zhou et al. 2014; Kuhl et al. 2021; Han et al. 2022; Niwa et al. 2025). Conversely, turnover may be more likely when sex determination remains developmentally flexible, acts over broader sensitive windows, integrates genetic, epigenetic, germ-cell, or environmental inputs, and lacks chromosome-wide dosage compensation. Such architectures may provide more entry points for new upstream regulators, allowing alternative sex-determining loci to arise and replace existing systems. We do not argue that developmental GRN architecture is the only factor shaping turnover. Processes such as tolerance to sex reversal, hybridization, gene duplication, and neofunctionalization after polyploidy may also promote sex chromosome turnover in particular lineages (Evans et al. 2024). However, these factors are not universal across ectothermic vertebrates and are not considered further in this review. Here, we focus on developmental GRN architecture as a broad and testable framework for explaining why some systems may be more permissive to sex chromosome replacement than others.

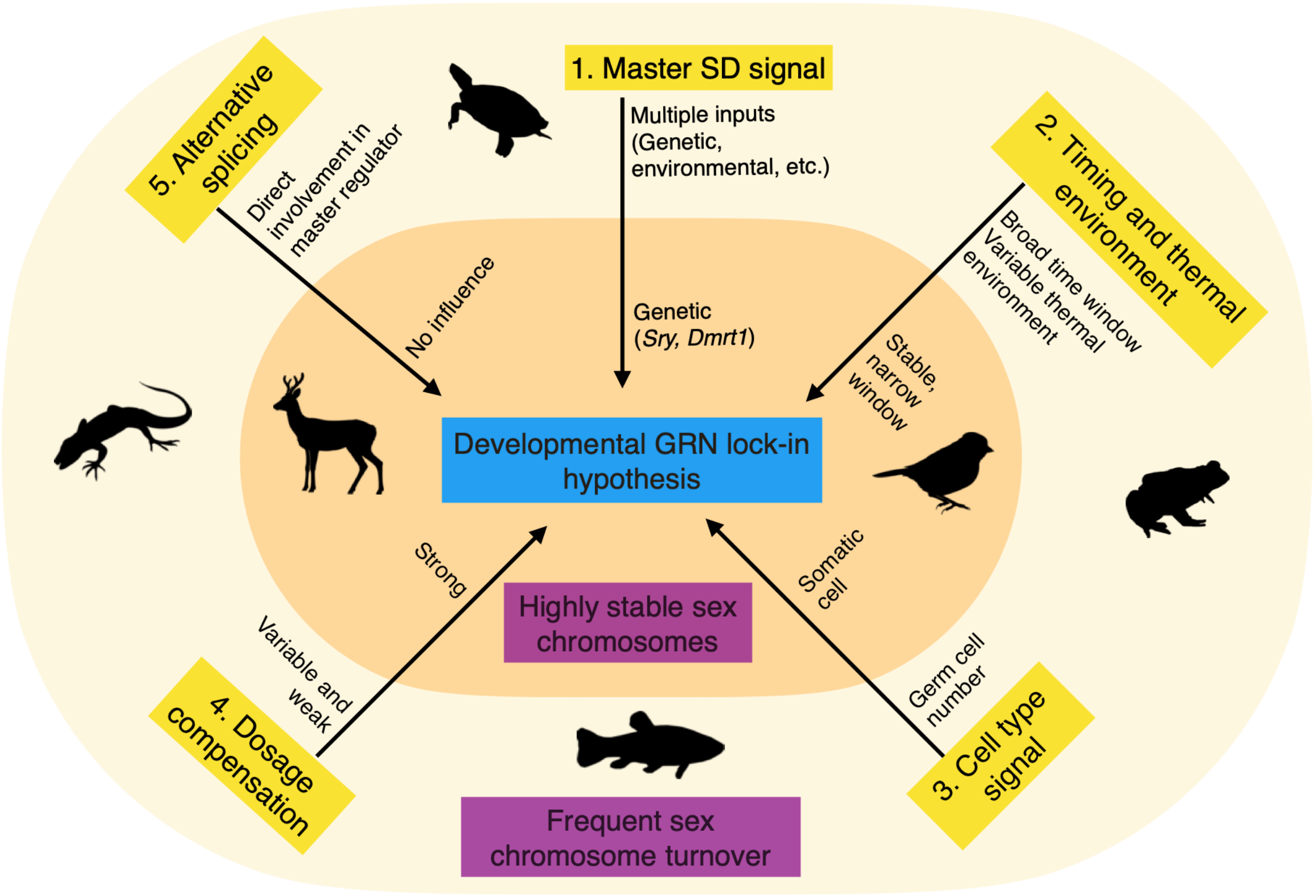

**Figure 2**. Conceptual framework of the developmental GRN lock-in hypothesis: illustration of multiple interacting variables that may affect developmental gene regulatory networks (GRNs) and sex chromosome dynamics across major vertebrate groups. Mammals and birds fall toward one end of the spectrum, characterized by a narrow embryonic window of sex determination, early somatic cell initiation, strictly genetic master switches, and strong regulatory lock-in. Many fishes, amphibians and non-avian reptiles fall toward the other end, where genetic and environmental inputs jointly affect the master sex-determining signals, sex determination occurs over a broader embryonic window, thermal conditions are less stable, a threshold germ cell number may be required, and alternative splicing acts closer to the decision point. Together, these features are associated with greater flexibility in sex-determining pathways and more frequent sex chromosome turnover. The diagram was created manually in Keynote, and the animal silhouettes were obtained from open-access PhyloPic resources: https://www.phylopic.org.

**Table 1.** Evolutionary hypotheses proposed to explain sex chromosome turnover or long-term stability. The hypotheses listed are not mutually exclusive and may interact. For example, the hot-potato model combines sexually antagonistic selection and mutation load accumulation hypotheses (Blaser et al. 2014).

| Hypothesis/ framework | Main mechanism | What it can explain | Main limitations | Citations |
|---|---|---|---|---|
| **Sexually antagonistic selection** | A new sex-determining system may spread if it becomes linked to alleles beneficial to one sex but deleterious to the other in another chromosome, thereby improving sex-specific fitness. | Can explain turnover when a new SD locus captures favourable sexually antagonistic variation. It can also explain non-random recruitment of genomic regions if some regions repeatedly harbour sexually antagonistic alleles. | Direct empirical evidence that sexually antagonistic alleles drive turnover remains limited. It does not easily explain repeated turnover cycles once sexually antagonistic alleles are already linked to an existing sex chromosome. | Van Doorn and Kirkpatrick 2007; Van Doorn and Kirkpatrick 2010; van Doorn 2014 |
| **Mutation load accumulation** | Suppressed recombination causes deleterious mutations to accumulate on the Y or W. A new sex-determining system can be favoured if it escapes the loaded sex-limited chromosome, promoting turnover. | Can explain repeated turnover while preserving the same heterogametic system, for example XY-to-XY or ZW-to-ZW transitions. | Requires sufficient mutation load on the sex-limited chromosome, which is rarely directly quantified. It may not explain turnovers in very young or weakly differentiated systems, and it does not by itself explain which new SD locus or chromosome is recruited. | Blaser et al. 2013 |
| **Evolutionary trap** | Once sex chromosomes become highly differentiated, degenerated and functionally specialized, replacing them becomes costly. Turnover is therefore suppressed because alternative genotypes would suffer from dosage imbalance, deleterious Y/W fixation, or sex-linked regulatory disruption. | Can explain long-term stability of ancient, highly differentiated systems such as therian XY and avian ZW systems. Also predicts lower turnover in heteromorphic and strongly dosage-compensated systems. | Differentiation/degeneration alone may be an incomplete proxy for the regulatory and developmental constraints that suppress turnover. Does not easily explain long-term stability of some homomorphic systems. | Pokorná & Kratochvíl, 2009 |
| **Meiotic drive / selfish genetic elements** | Sex-ratio distortion caused by meiotic drive, endosymbionts, hybridization, or other sex-biased processes can favour a new sex-determining allele that restores a more balanced sex ratio. | Can explain rapid turnover when populations experience biased sex ratios. It can also explain changes in heterogamety, because the direction of transition depends on whether the distortion favours males or females | Requires evidence for sex-ratio distortion, drive, endosymbionts, or hybridization. Such evidence is available only in some systems. It is unlikely to explain turnover broadly across all fishes, amphibians, and reptiles. | Werren and Beukeboom 1998; Kozielska et al. 2010; Ubeda et al. 2015 |
| **Genetic drift** | New sex-determining alleles can move through genotype-frequency paths that maintain an | Can explain turnover without requiring additional forces like sexually antagonistic | Mostly theoretical and difficult to distinguish empirically from weak | Bull and Charnov 1977; |

| Hypothesis/ framework | Main mechanism | What it can explain | Main limitations | Citations |
|---|---|---|---|---|
| | approximately balanced sex ratio. In small populations, drift may allow a new sex chromosome system to replace the ancestral one without strong directional selection. | alleles, mutation load, or sex-ratio distortion. Also predicts that heterogamety may often be preserved because X or Z chromosomes are initially more frequent than Y or W chromosomes. | selection. Does not explain why some chromosomes or genes are repeatedly recruited, or why turnover is especially common in some lineages but rare in others. | Veller et al. 2017; Saunders et al. 2018 |
| **Chromosome-specific drift under stabilizing selection** | Under stabilizing selection on polygenic traits, chromosomes can accumulate chromosome-specific contributions to complex traits. Novel sex-determining genotypes produced during turnover may then have reduced fitness, creating a barrier to turnover and favouring stability. | Can explain long-term stability without requiring sexually antagonistic alleles or Y/W degeneration. It also predicts that turnover should be constrained by genome-wide polygenic architecture, especially for sexually dimorphic traits. | Currently mainly theoretical. It does not directly specify developmental mechanisms, cell types, dosage regulation, or which GRN nodes can become new SD loci. Comparative empirical tests are still needed. | Muralidhar 2026 |
| **Developmental GRN lock-in** | Turnover is constrained when sex determination is embedded in an early, canalized developmental GRN and coupled to strong dosage regulation. A new SD allele must not only spread genetically, but also function at the correct developmental time, in the correct cell type, and within the existing dosage-regulatory context. | Can explain why turnover is rare in lineages with early, somatic, environmentally buffered master switches and strong dosage regulation, but more frequent in lineages with broader sensitive windows, environmental responsiveness, germ-cell involvement, alternative splicing, and weaker or local dosage regulation. | Proposed here as a testable framework rather than a statistically established explanation. Broad comparative tests require developmental, genomic, and dosage-compensation data that are not yet available for many vertebrate lineages. | This review |

## 2. Developmental gene regulatory networks: a neglected dimension of sex chromosome evolution

Much of our understanding on the genetics of sex determination in vertebrates has been shaped by the mammalian model, in which sex determination is often treated as synonymous to the action of a single MSD. By contrast, work in teleost fishes, amphibians and reptiles shows that sex determination can involve multiple genetic, epigenetic and environmental inputs, acting at different positions and times within the GRN. These differences in how the network is wired and regulated may strongly influence how easily new sex-determining loci can be integrated and, consequently, how readily sex chromosomes can be replaced.

In this review, we synthesise developmental and dosage regulatory evidence across vertebrates and propose a multidimensional framework in which stable versus labile sex chromosome systems differ along several interacting axes (Figure 2): (i) the master sex-determining signal (strictly genetic and cell-autonomous versus integrative signals combining genetic, epigenetic and environmental inputs); (ii) the sensitive period for sex determination and its interaction with the thermal environment (narrow early window in a homeostatically buffered embryo versus a prolonged, temperature-sensitive period); (iii) the cell type of the MSD is initially expressed in (somatic cell with master genetic switches versus germline cell number thresholds); (iv) gene expression regulation such as dosage compensation (chromosome-wide mechanism versus absence or gene specific mode); (v) the influence of regulatory plasticity, such as environmentally responsive alternative splicing that acts within the SD cascade (little role versus acting near key node affecting sex determination). We suggest that mammals and birds occupy one end of this spectrum, whereas many teleost fishes, amphibians and non-avian reptiles occupy the other. Our aim is to complement evolutionary/population genetic models of sex-chromosome evolution with a developmental perspective: developmental GRN architecture is shaped by and constrains sex chromosome evolution, creating feedback loops that can lead either to long-term evolutionary "lock-in" or to recurrent turnover. In the sections that follow, we develop this framework in details and discuss how integrating developmental biology, molecular mechanism, evolutionary genomics and population genetics can refine our understanding of sex-chromosome turnover across vertebrates.

## 3. Master sex-determining genes acting on somatic cells and endothermy ensures developmental canalization of sex determination in mammals and birds

In therian mammals, sex determination is classically framed as the bifurcation of the bipotential gonad into testis or ovary, with most downstream sexual dimorphism caused by hormones produced by the gonads rather than from direct genetic effects in other tissues (Kashimada and Koopman 2010; Mäkelä et al. 2019). The Y-linked gene *Sry* acts as the MSD in almost all examined therians, with only a handful of exceptions (Gubbay et al. 1990; Koopman et al. 1991; Kashimada and Koopman 2010; Terao et al. 2022). In mouse, *Sry* expression is both cell-type specific and sharply time-limited during early embryonic development. Reporter and expression analyses show that *Sry* is confined to somatic pre-Sertoli cells of the genital ridge and is detectable only during a transient interval from approximately embryonic day E10.5 to E12.5 (Albrecht and Eicher 2001; Bullejos and Koopman 2001). Within this narrow window, *Sry* directly activates *Sox9* via testis-specific enhancers and triggers a positive-feedback loop in which *Sox9* maintains its own expression and reinforces testis fate, while mutually antagonising pro-ovarian pathways such as *Wnt4*/*Rspo1*/*β-catenin* (Sekido and Lovell-Badge 2008; Kashimada and Koopman 2010; Herpin and Schartl 2015). Single-cell transcriptomics analyses further refine the picture: during foetal testis development, *Sry* is transiently activated specifically in pre-Sertoli cells and its onset directly precedes their differentiation into Sertoli cells with high *Sox9* expression. *Sry* expression sweeps along the genital ridge in a centre-to-pole wave, with individual supporting pre-Sertoli cells passing through a brief *Sry*$^{+}$/Sox9 low state before switching to a stable *Sox9*-dominated identity, consistent with *Sry* acting as a short-lived trigger that hands over control to a self-sustaining *Sox9* network (Koopman 1999; Stévant et al. 2019; Stévant and Nef, 2019). Disruption of *Sry*, or of key *Sox9* regulatory elements, in XY embryos results in loss or failure of *Sox9* up-regulation and a robust male-to-female gonadal sex reversal, with supporting cells adopting the female development pathway such as granulosa-like identity and ovaries forming instead of testes (Koopman et al. 1991; Lavery et al. 2011; Gonen et al. 2018). Furthermore, germ-cell ablation experiments in mice and humans indicate that germ cells are not required for the initial specification of testis versus ovary: Sertoli cells, testis cords and ovarian structures can form and steroidogenesis can be initiated in their absence (Maatouk et al. 2012; Mäkelä et al. 2019). Germ cells have a secondary role, contributing to optimal cord morphogenesis and endocrine function of the foetal testis, after the sex determination pathway has been initiated (Liu et al. 2010; Mäkelä et al. 2019). Thus, in mammals, the primary fate decision is clearly made by somatic cells and driven by *Sry* activating *Sox9*, with germ cells acting later as modulators rather than initiators of gonadal sex determination.

In birds, dosage of the Z-linked transcription factor *Dmrt1* plays an analogous role to *Sry* in mammals. The exact mechanism is different as in birds the males are the homogametic sex (ZZ males), and *Dmrt1* is expressed in both sexes. Across major avian clades ZZ embryos have roughly twice the *Dmrt1* expression of ZW embryos in the bipotential gonad before gonadal sexual differentiation, which is consistently associated with testis development (Graves 2014; Hirst et al. 2017). Knockdown or knockout of *Dmrt1* in chicken demonstrate that ZZ embryos induce gonadal feminisation, with ovarian-like morphology, up-regulation of ovarian markers such as *Foxl2* and *Aromatase*, and down-regulation of testis markers including *Sox9* (Smith et al. 2009; Ioannidis et al. 2021; Lee et al. 2021). Nevertheless, these feminised ZZ gonads do not become fully functional ovaries, and adult *Dmrt1*-disrupted ZZ birds retain male secondary sexual traits (Lee et al. 2021). Overall, current evidence supports *Dmrt1* dosage as the primary genetic switch for testis development in birds, acting largely cell-autonomously in gonadal somatic cells.

Mammals and birds share several developmental GRN features relevant to sex-chromosome stability. First, their primary signals are strictly genetic and chromosomally encoded without any environmental input at the time of cell fate specification. In both clades, the initial bias toward testis development depends on a discrete chromosomal cue, *Sry* expression (Y-linked) on XY individuals, or high (Z-linked) *Dmrt1* dosage on ZZ individuals (Manolakou et al. 2006; Barske and Capel 2008; Kashimada and Koopman 2010). Environmental and endocrine factors can certainly disrupt sexual development, but available data suggest that they normally act downstream of the primary switch, both in the sex determination developmental GRN and in later development (Manolakou et al. 2006; Barske and Capel 2008). Second, after a short time window, altering *Sry* or *Dmrt1* expression no longer affects cell fate, because downstream positive and negative feedback loops within the GRN lock the gonad into a testis or ovary trajectory (Sekido and Lovell-Badge 2008; Sekido and Lovell-Badge 2009; Kashimada and Koopman 2010; Mäkelä et al. 2019). Third, *Sox9* activation by *Sry* in mammals and *Dmrt1* dosage in birds occurs within strongly canalised developmental GRN, that is, networks buffered against stochastic noise and perturbations so that, once induced, they reliably drive the gonad along a testis trajectory. These networks coordinate differentiation of multiple somatic lineages (Sertoli, Leydig, granulosa and the gonadal vasculature) and link gonadal fate to various endocrine pathways (Kashimada and Koopman 2010; Herpin and Schartl 2015; Mäkelä et al. 2019). This developmental entrenchment makes it difficult for alternative loci on other chromosomes to take over as primary sex-determining switches without causing widespread pleiotropic damage. For example, disruption at this level is typically associated with disorders of sexual development and infertility in both humans and mouse models, and karyotypic deviations from XX♀/XY♂ or ZW♀/ZZ♂ usually generate severe developmental abnormalities (Warr and Greenfield 2012; Croft et al. 2018; Mäkelä et al. 2019).

The canalisation of developmental GRN is probably further reinforced by endothermy. Mammals and birds generally maintain embryos within a tightly regulated thermal range, insulating the sex-determining cascade from ambient temperature fluctuations (Barske and Capel 2008; Luckenbach et al. 2026). Comparative and macroevolutionary studies have repeatedly noted an association between endothermy, predominance of genotypic sex determination, and the presence of highly differentiated, evolutionarily stable sex chromosomes in mammals and birds, in contrast to the more labile systems typical of many ectothermic vertebrates (Miller et al. 2004; Barske and Capel 2008; Rovatsos et al. 2017; Kratochvíl, Stöck, et al. 2021). From a macroevolutionary perspective, therian XY and avian ZW systems represent two of the oldest and best-characterized conserved sex-chromosome systems in amniotes, though ancient and stable sex-chromosome systems have also been described in other vertebrate lineages, including certain fishes. Comparative genomic analyses suggest that the ancestral therian XY pair originated around 160–180 million years ago, whereas the ancestral avian ZW system is at least 60–100 million years old (Graves 2006; Ohno 2010; Zhou et al. 2014; Hughes et al. 2015; Xu et al. 2019; Kuhl et al. 2021; Niwa et al. 2025). Across both clades, the X and Z chromosomes retain many ancestral genes, while the core developmental GRN has persisted for tens of millions of years despite extensive lineage-specific degeneration. Although sex chromosome-autosome fusions have generated neo-sex chromosomes in several mammalian lineages, these events generally extend the sex-linked region of the ancestral X or Y chromosome rather than replacing the primary sex-determining locus (Ohno 2010; Graves 2014; Sigeman et al. 2019; Kratochvíl, et al. 2021).

In summary, the combination of (i) an early-acting, transient, strictly genetic master signal expressed in somatic supporting cells, (ii) strong positive and negative feedback loops on gene expression canalising testis (Sertoli) vs over (granulosa) developmental trajectories, (iii) thermal buffering of embryonic development through endothermy, and possible long-term coevolution of these developmental networks with highly differentiated X/Y or Z/W chromosomes, together creates a developmental GRN lock-in that strongly resists invasion by alternative sex-determining loci. Together these effects make successful invasion and thus sex-chromosome turnover extremely unlikely (Herpin and Schartl 2015; Hughes et al. 2015; Charlesworth 2021b; Kratochvíl, Gamble, et al. 2021). This provides a clear contrast with the more flexible, environmentally integrated sex-determining GRNs of many ectothermic vertebrates, which we consider in the next section.

## 4. Flexible, integrative sex-determining GRNs in amphibians, teleost fishes and non-avian reptiles

Many amphibians, teleost fishes and non-avian reptiles exhibit more flexible developmental GRNs than mammals and birds, in which genetic, environmental and germ-cell-derived signals are integrated over a broad developmental window to determine sex. These clades harbour much of the known diversity in sex-determining mechanisms and sex chromosome systems, including environmentally-dependent sex determination (ESD), which may interact with genetic sex determination (GSD), transitions between the two, and rapid sex-chromosome turnover (Kobayashi et al. 2013; Pan et al. 2016; Nagahama et al. 2021; Pan et al. 2021; Uno and Matsubara 2024; Augstenová and Ma 2025).

Across amphibians, although GSD appears to be common, the sex chromosomes and MSDs have been identified in only a small number of lineages, where they are scattered phylogenetically and vary substantially among groups. Linkage mapping and comparative analyses show that amphibian sex chromosomes derive from multiple, non-homologous linkage groups, even within single families, rather than from one conserved ancestral pair (Sumida and Nishioka 2000; Dufresnes et al. 2015; Jeffries et al. 2018; Ma and Veltsos 2021; Uno and Matsubara 2024). The W-linked *Dmrt1* paralogue *Dm-w* in *Xenopus laevis* is one of the best experimentally characterised MSD. It acts as a dominant female-determining gene, antagonises *Dmrt1*, and promotes ovarian development (Yoshimoto et al. 2008; Yoshimoto et al. 2010). More recently, *Bod1l* in *Bufo(tes) viridis* (Kuhl et al. 2024), *Dmrt1* in *Rana temporaria* (Carpentier et al. 2026), and *Sox3* in *Glandirana rugosa* (Miura et al. 2024) have also been identified or implicated as candidate MSDs, although their level of molecular functional validation differs among systems. However, environmental and endocrine cues may still play important roles in sex determination (Wallace et al. 1999; Ggert 2004; Nakamura 2009). Classical work and more recent experiments show that larval exposure to exogenous estrogens or androgens, or to altered temperatures, can bias sex ratios or cause complete sex reversal in several anurans, including *X. laevis* and the common European frog, *Rana temporaria* (Matsuba et al. 2008; Flament 2016; Ruiz-Garciá et al. 2021). A recent review concludes that no amphibian species is currently known with pure ESD, but many show significant thermal sensitivity superimposed on GSD including *Rana temporaria, R. japonica, Bufo bufo, Anaxyrus terrestris* and several others (Ruiz-Garciá et al. 2021; Akashi et al. 2024). Taken together, these data suggest that amphibian gonads are specified by GRNs modulated both by genes and by external factors.

Teleosts exemplify highly labile, integrative sex-determining networks. Studies have identified a wide range of MSDs including *Dmrt1/Dmy*, *Amh/Amhy*, *Amhr2*, *Gsdf*, *Sox3*, *Gdf6* and others, often residing on different chromosomes or linkage groups in closely related species (Guiguen et al. 2018; Pan et al. 2021; Kitano et al. 2024; Wang et al. 2024). This diversity underscores that the top of the sex determination GRN is easily rewired in teleosts. A distinctive feature of many teleost systems is that germ cell number and state are themselves decisive inputs into gonadal fate. For example, in zebrafish and stickleback, genetic or experimental depletion of primordial germ cells leads to phenotypic males containing testis-like gonads, whereas individuals retaining sufficient primordial germ cells can develop as females, and partial primordial germ cell depletion biases sex ratios towards males (Dai et al. 2015; Tzung et al. 2015; Treaster et al. 2025). In medaka, germ-cell-depleted XX embryos develop into males, whereas increasing germ cell numbers can feminize XY individuals (Kurokawa et al. 2007). These findings support a model in which germ cell number and proliferation, induced by genes such as *Dmrt1/Dmy*, *Amh/Amhy*, *Amhr2*, *Gsdf*, *Sox3*, *Gdf6*, determine whether the primordial, bipotential gonad crosses a male or female-promoting threshold, and the decision period is wider than in mammals and birds, whose decision is short and initiated by somatic cells.

Environmental factors frequently feed into the sex determination developmental GRN. Reviews of teleost sex determination consistently emphasize that genetic, endocrine, social and thermal cues converge on a network of gonadal regulators, with species-specific combinations of MSDs at the top (Baroiller et al. 2009; Kobayashi et al. 2013; Nagahama et al. 2021; Augstenová and Ma 2025). In several teleosts, temperature manipulation during early gonadal development can induce sex reversal and alter the expression and epigenetic state of key sex-determining genes. For example, in Nile tilapia, high-temperature exposure masculinizes genetic females and is associated with increased DNA methylation and reduced expression of the ovarian aromatase gene *Cyp19a1a* (Li et al. 2014; Wang et al. 2017). In European sea bass and related species, both the genetic background and rearing temperature influence methylation or expression of *Cyp19a1a* and *Dmrt1* and adult sex ratios (Piferrer and Blázquez 2005; Navarro-Martín et al. 2011).

Non-avian reptiles display the full continuum from purely GSD to ESD, as well as multiple evolutionary transitions from ESD to GSD systems (Valenzuela and Lance 2004; Ezaz et al. 2010; Holleley et al. 2015). Although our focus is not on ESD itself, temperature-dependent SD (TSD) provides important insight into how flexible developmental GRNs can translate environmental cues into gonadal fate. In classical temperature-dependent sex determination (TSD) species, including the red-eared slider turtle (*Trachemys scripta*), many freshwater turtles, sea turtles and crocodilians, incubation temperature during a prolonged thermosensitive period biases gonadal development (Valenzuela and Lance 2004; Akashi et al. 2024). In *T. scripta*, the molecular basis of TSD has begun to be elucidated: the histone demethylase *Kdm6b* shows temperature-dependent, sexually dimorphic expression in early embryos, and loss-of-function experiments demonstrate that it is required for testis development at male-producing temperatures (Venegas et al. 2016; Ge et al. 2018; Weber et al. 2020; Martínez-Pacheco et al. 2024). Consistent with a prolonged thermosensitive window, transcriptomic and epigenetic studies in turtles, including sea turtles, show that sex-biased gene expression, DNA methylation, chromatin state, and other regulatory differences accumulate gradually across this interval (Venegas et al. 2016; Ge et al. 2018; Whiteley et al. 2021; Akashi et al. 2024; Martínez-Pacheco et al. 2024).

Mixed GSD-TSD systems further highlight how environmental input can interact with, and sometimes override, chromosomal sex determination. In the Australian central bearded dragon *Pogona vitticeps* (ZZ/ZW), high incubation temperatures cause male-to-female sex reversal of ZZ embryos, and the resulting sex-reversed ZZ females are fertile and produce all-ZZ offspring, providing a clear route by which W chromosomes could be lost or replaced over one generation (Holleley et al. 2015). Transcriptomic analyses using sex reversed individuals show extensive rewiring of gonadal GRNs, including changes in *Dmrt1*, *Foxl2*, *Cryp19a1a* and other core pathway genes, as well as coordinated changes in somatic and germ cell chromatin state and marked differences in meiotic timing between developing testes and ovaries (Whiteley et al. 2021). More broadly, data from amphibians and reptiles indicate that temperature and endocrine cues can re-specify gonadal fate during larval or embryonic development, often via effects on epigenetic regulators and splicing factors within the gonadal GRN (Ge et al. 2018; Ruiz-Garciá et al. 2021; Sun et al. 2025).

Taken together, these observations support an emerging picture of developmental GRNs that are more permissive to change in amphibians, teleosts and non-avian reptiles than those of mammals and birds. The reason is that the sex determination decision is produced by the interaction of multiple genetic, epigenetic and environmental inputs, including the internal environment such as germ cell number or the endocrinal environment which take place over an extended and environmentally exposed decisive period. Consequently, there are multiple points at which new sex-determining alleles, environmental modifiers can be accommodated without catastrophic disruption of development. This GRN architecture is consistent with, and may contribute to, the high lability of sex determination and frequent sex-chromosome turnover documented in many amphibian, teleost and non-avian reptile lineages, even though direct causal links between specific GRN features and particular sex-chromosome transitions remain to be demonstrated.

## 5. Dosage compensation as a developmental constraint

Classical models of sex-chromosome evolution predict that suppression of recombination around the sex-determining locus promotes gene loss on the Y or W and generates dosage imbalances that are often offset by evolution of dosage compensation in the heterogametic sex (Charlesworth and Crow 1978; Charlesworth 1998; Mank 2013; Graves 2015). The strength, mechanism and developmental timing of dosage compensation differ widely among lineages and are now routinely summarized using sex-chromosome: autosome expression ratios and analyses of sex-biased expression (Naurin et al. 2010; Mank 2013; Duan and Larschan 2019). From a developmental GRN perspective, dosage compensation may do more than correct sex-linked dosage imbalance. Once established, it creates a sex-chromosome-specific regulatory environment that operates early in development, before or alongside the onset of gonadal sex determination in mammals and birds. This early regulatory layer may interact with the timing and cell-type specificity of MSDs, making later changes in sex-determining architecture more constrained and potentially reinforcing further differentiation between sex chromosomes (Lenormand et al. 2020; Lenormand and Roze 2022; Lenormand and Roze 2025).

In therian mammals, dosage compensation is achieved primarily through X-chromosome inactivation (XCI) in XX individuals. Dosage balance is shaped by up-regulation of X-linked gene expression from the active X, although XCI is incomplete and a subset of genes escapes silencing on the inactive X (Lin et al. 2012; Carrel and Brown 2017; Sauteraud et al. 2021; Lister et al. 2024; San Roman et al. 2024). Interestingly, the incomplete dosage balance at transcriptional level is balanced at the protein level by post transcriptional compensation in platypus and chicken (Lister et al. 2024). Importantly, mammalian dosage regulation is also shaped by the evolutionary fate of Y-linked genes. Although most ancestral Y-linked genes were lost during Y chromosome degeneration, the surviving genes were not retained randomly: retained Y gametologs are enriched for broadly expressed, dosage-sensitive regulators, and many of their X-linked homologues escape XCI, suggesting that X–Y gene pairs contribute to dosage balance between sexes (Bellott et al. 2014; Martínez-Pacheco et al. 2020). At the same time, some Y-linked genes have acquired or retained testis-biased expression, linking Y chromosome evolution to male reproductive function (Bellott et al. 2014; Martínez-Pacheco et al. 2020).

Developmental studies in mouse show that XCI imprinting is initiated in extra-embryonic tissues at the pre-implantation stage, followed by reactivation of the paternal X in the epiblast and onset of random XCI during early development (~E4.5–E6.5), well before gonad formation (Kamikawa and Donohoe 2014; Oikawa et al. 2014; Malcore and Kalantry 2024). By the time the genital ridges emerge (~E10) and the MSD (*Sry*) is transiently expressed in pre-Sertoli cells (~E10.5–E12.5), a stable pattern of single random active X chromosome is already in place in somatic cell lineages (Oikawa et al. 2014; Panda et al. 2020). The mammalian sex chromosomes are shaped by several distinct regulatory processes acting at different developmental stages. During embryonic testis determination, transient expression of the Y-linked *Sry* gene activates *Sox9* and initiates the male developmental pathway. Separately, XCI regulates X-linked gene dosage in females, whereas during male meiosis, meiotic sex chromosome inactivation (MSCI) transcriptionally silences much of the X and Y chromosomes in spermatocytes (Turner 2007; Okamoto et al. 2011). These stage-specific regulatory mechanisms make the mammalian sex chromosomes epigenetically distinct from autosomes. At the same time, extensive Y-chromosome degeneration has left a reduced set of genes enriched for dosage-sensitive and male-reproductive functions (Hughes and Page 2015). In this context, relocating the MSD from the Y to an autosome, either by *Sry* translocation or by recruitment of a novel autosomal gene, would require the new locus to function within a very narrow embryonic window and within an already established dosage-regulatory regime. Immediate genotypes, such as XX individuals carrying an autosomal *Sry*, would still follow the XCI program and dosage compensation, creating mismatches between gonadal sex and X-linked dosage regulation, and Y-linked gene function. The combination of (i) a pre-established chromosome-wide dosage-compensation programme, (ii) retained dosage-sensitive X-Y gametologs, and (ii) a tightly timed somatic cell *Sry* trigger therefore creates a form of developmental GRN lock-in: the sex-determining cascade is anchored to the X/Y pair, and alternative top-level configurations are likely to generate pervasive dosage and expression mismatches (Hughes and Page 2015; Charlesworth 2021a).

In birds, dosage compensation of the Z chromosome is incomplete and heterogeneous. Analyses consistently show that many Z-linked genes remain male-biased (ZZ > ZW) and that female Z expression is not fully equalised to the male Z, arguing against a mammal-like chromosome-wide inactivation mechanism (Mank 2009; Uebbing et al. 2015; Sigeman et al. 2018; Fallahshahroudi et al. 2025). Instead, studies support a model of gene-by-gene and tissue-specific dosage adjustment: some Z-linked loci show equal expression between sexes in particular tissues or stages, whereas many others remain male-biased, and epigenetic marks on the female Z are consistent with local modulation rather than global silencing (McQueen and Clinton 2009; Naurin et al. 2010; Sigeman et al. 2018). For sex determination, the key observation is that *Dmrt1* appears to escape full compensation in the embryonic gonad. During the bipotential stage, *Dmrt1* expression is roughly double in ZZ gonads compared to ZW gonads, and experimental reduction of *Dmrt1* dosage in ZZ embryos via RNAi or CRISPR/Cas9 knock-out causes gonadal feminization with ovarian-like morphology, up-regulation of ovarian markers and down-regulation of *Sox9* and other testis-developmental related genes (Smith et al. 2009; Lambeth et al. 2014; Ioannidis et al. 2021). In other tissues, however, *Dmrt1* and many Z-linked genes have roughly equal expression between the sexes (Uebbing et al. 2013; Estermann et al. 2021). From a GRN standpoint, this means that the avian MSD operates in a chromosomal context that is partially dosage-regulated depending on the genes in question. There is enough dosage compensation to mitigate deleterious Z-linked dosage effects globally, but the *Dmrt1* dosage difference is preserved specifically in the cell types and time window relevant to sex determination (Smith et al. 2009; Lambeth et al. 2014; Ioannidis et al. 2021). The Z-linked regulatory environment is distinct from autosomes but less rigid than the XCI landscape in mammals. In addition, meiotic sex chromosome inactivation also evolves the developmental rewiring of gene expression silencing in Z-linked genes in female germline cells during meiosis (Schoenmakers et al. 2009), which contributes to developmental GRN lock-in.

In vertebrate groups outside mammals and birds, dosage-compensation mechanisms are more variable across species and lineages and, when they evolve, are typically localized or gene-specific rather than chromosome-wide. In amphibians, no dosage compensation mechanism of any kind has been demonstrated. Comparative expression studies in frogs indicate that males and females can differ in the copy number of sex-linked genes without corresponding global equalization of expression, implying that these lineages often tolerate dosage differences and do not need to balance them using chromosome-scale mechanisms (Malcom et al. 2014). Many teleost fishes show a similar pattern. Across most surveyed species with largely homomorphic sex chromosomes, the gene expression ratios between sex chromosomes and autosomes are close to 1 in both sexes, suggesting either an absence of chromosome-wide compensation or compensation restricted to a subset of dosage-sensitive genes (Mank 2013; Duan and Larschan 2019). Some notable exceptions are *Poecilia picta*, *P. parae* and *P. bifurca* where massive gene losses were detected along the highly diverged Y chromosome, and there is nearly complete dosage compensation of X-linked genes in both sexes (Darolti et al. 2019; Fong et al. 2023; Fong et al. 2025). However, such cases remain rare among teleosts; in many systems, sex chromosomes are largely homomorphic and retain an “autosome-like” expression profile for much of their gene content, even when some degree of degeneration has occurred. Non-avian reptiles occupy an intermediate position. Work on lizards and turtles has revealed incomplete or regional Z dosage compensation in several ZW and XY systems: some Z-linked genes are dosage-balanced between sexes, whereas many remain male-biased (Zhu et al. 2022; Webster et al. 2023). In the Gila monster (ZZ/ZW), for example, global Z expression is higher in males than females, but there is evidence for local gene-by-gene compensation of a subset of Z-linked genes (Webster et al. 2023). An exception showing chromosome-level dosage compensation in the green anole (*Anolis carolinensis*) which has a differentiated XY system (Tenorio et al. 2024). Together, amphibians, most teleosts and many non-avian reptiles have sex chromosomes without chromosome-wide dosage compensation but sometimes have dosage compensation for dosage sensitive genes (Mank 2013; Gu and Walters 2017; Rovatsos and Kratochvíl 2021) the context of developmental GRN lock-in hypothesis, the chromosome-wide dosage compensation, tuned to the ancestral XX/XY or ZZ/ZW system of mammals and birds, makes sex chromosome turnover difficult. A novel SD locus would allow the generation of XX males, XY females (or ZZ females, ZW males) with inappropriate X or Z dosage, resulting in deleterious genotypes that would block sex chromosome turnover.

Therian mammals represent a strong form of developmental GRN lock-in: X-chromosome inactivation is established early in development, the *Sry→Sox9* switch acts within a narrow developmental window in pre-Sertoli somatic cells, and the sex chromosomes are regulated by chromosome-wide dosage compensation (Ohno 1966; Koopman et al. 1991; Sekido and Lovell-Badge 2009; Graves 2016). Birds may represent a partial lock-in of the primary sex-determining mechanism: the ancient ZW system and dosage-sensitive role of *DMRT1* are deeply integrated into gonadal regulatory networks (Smith et al. 2009). Z-chromosome dosage compensation, however, remains incomplete and gene-specific (Itoh et al. 2011; Zimmer et al. 2016; Sigeman et al. 2018), indicating a regulatory architecture distinct from the chromosome-wide dosage compensation characteristic of mammals. This difference may influence how structural changes are accommodated, but it does not by itself imply greater structural lability, since neo-sex chromosomes can also evolve in mammals through mechanisms that preserve expression and meiotic pairing. We therefore view avian 'lock-in' primarily as conservation of the core sex-determining mechanism rather than complete structural constraint. In contrast, many ectothermic vertebrates with labile sex determination and frequent sex-chromosome turnover lack chromosome-wide dosage compensation and often retain sex chromosomes with indistinguishable chromatin and gene expression to autosomes. In such systems, new MSDs may be more able to evolve, move between chromosomes, and establish in populations with fewer conflicts with pre-existing strong dosage regulatory programmes.

We emphasise that current data do not support a simple causal link between dosage compensation and sex-chromosome stability: there are lineages with old, stable sex chromosomes but only local dosage compensation, and lineages with strong dosage compensation with sex chromosome fusions, rearrangements or turnover (Mank 2009; McQueen and Clinton 2009; Mank 2013; Sigeman et al. 2018; Furman et al. 2020), as well as rare cases of sex chromosome turnover in mammals (Schartl and Herpin 2018; Terao et al. 2022). The value of the developmental GRN lock-in perspective is that it integrates dosage compensation with other features discussed in this review into a coherent regulatory network framework.

## 6. Alternative splicing and gene regulatory plasticity in vertebrate sex determination

Alternative splicing adds a post-transcriptional layer of flexibility to the developmental GRN framework. In vertebrates, many core regulators of gonadal development, including *Wt1, Sox9, Foxl2, Amh, Fgfr2* and *Dmrt1*, produce multiple isoforms with distinct DNA-binding properties, transactivation potential or interaction partners (Hammes et al. 2001; Gómez-Redondo et al. 2021). Genome-wide transcriptomic analyses of mouse foetal gonads (across E10.5–E13.5) identify hundreds of sex- and stage-specific splicing events, enriched in genes involved in transcription, chromatin regulation and RNA processing (Zhao et al. 2018; Gómez-Redondo et al. 2021). These data suggest that alternative splicing is an important regulatory layer within developmental GRNs that control sexual fate, that can act both near the sex-determining decision point or during downstream gonadal sex differentiation, depending on the organisms (Augstenová and Ma 2025).

In therian mammals, alternative spicing matters for gonadal development, but current evidence suggests it does not affect the transcriptionally defined master switch for sex determination. The decisive male-determining event remains the tightly timed transcriptional activation of *Sry* and subsequent up-regulation of *Sox9* in somatic supporting cells, followed by strong positive feedback and mutual antagonism with ovarian gene expression. To date, there is no convincing evidence that splicing of *Sry* itself, or of *Dmrt1* in birds, is environment-sensitive or that it constitutes the primary male–female switch (Gómez-Redondo et al. 2021). Nevertheless, a specific *Fgfr2c* isoform mediates *Fgf9* signalling in Sertoli cell precursors and represses the *Foxl2/Wnt4*-driven ovarian pathways; disruption of *Fgfr2c* leads to partial or complete gonadal sex reversal in XY mice (Kim et al. 2007; Bagheri-Fam et al. 2017; Bird et al. 2023). *Foxl2* itself shows isoform diversity and is crucial for ovarian differentiation and maintenance, and its dysfunction causes ovarian failure and adult ovary-to-testis differentiation (Georges et al. 2013; Georges et al. 2014). Mis-splicing of several of these genes is implicated in disorders of sex development (Gómez-Redondo et al. 2021). RNA-seq studies confirm extensive alternative splicing in foetal mouse gonads exactly during the period when the bipotential gonad commits to a testis or ovary fate (Zhao et al. 2018). In birds, *Dmrt1* also produces multiple transcripts, but experimental work strongly supports *Dmrt1* dosage, rather than alternative isoforms, as the key determinant of testis development. Knockdown of *Dmrt1* in ZZ chicken embryos feminises the gonads (Smith et al. 2009), and combined genomic and functional analyses show that *Dmrt1* dosage is essential for primary sex determination (Estermann et al. 2021; Ioannidis et al. 2021). Together with early X-inactivation in mammals and locus-specific Z dosage adjustment in birds, these findings suggest that in endotherms alternative splicing operates largely within a pre-existing, dosage-compensated, somatic master-switch architecture. Changes in splicing at key nodes can strongly disrupt gonad development, but there is little evidence that they have directly contributed to sex-determining switches. This pattern is consistent with the long-term stability of XY and ZW systems and the rarity of master-switch replacement in these clades.

In ectothermic vertebrates with labile sex determination, alternative splicing often lies closer to the decision node for sex-determining switches and can often be directly modulated by environmental cues. Reptiles with TSD provide the clearest mechanistic examples. In the red-eared slider turtle (*T. scripta*), incubation temperature alters splicing of chromatin regulators such as *Kdm6b* and *Jarid2*; *pStat3–Kdm6b* signalling has previously been shown to couple temperature to *Dmrt1* activation in this system (Ge et al. 2018; Weber et al. 2020). Most recently, the RNA-binding protein RBM20 was shown to mediate TSD in this species, by modulating the relative abundance of two alternative isoforms of *Wt1* (+KTS and −KTS); perturbing RBM20 or *Wt1* splicing is sufficient to shift gonadal fate, establishing a direct causal link between temperature, splicing regulation and vertebrate sex determination (Sun et al. 2025). The dragon lizard *Pogona vitticeps* shows sex reversal of ZZ males at high temperature. At the molecular level, high temperatures alter alternative splicing of chromatin regulators: intron-retained and truncated isoforms of *Jarid2*, *Kdm6b* and *Jmjd3* accumulate specifically in sex-reversed females, implicating splice-form shifts in the TSD override of GSD (Deveson et al. 2017; Whiteley et al. 2022). These data again place temperature-sensitive alternative splicing of chromatin regulators at the interface between environment and the sex-determining GRN. In teleost fishes, multiple isoforms of *Dmrt1* and *Cyp19a1a* have been described, and isoform usage shifts during sex change or thermal sex reversal. In the protandrous hermaphrodite barramundi (*Lates calcarifer*), sex-specific methylation and alternative splicing of *Dmrt1* and *Cyp19a1a* are associated with male-to-female sex change: males lack full-length *Cyp19a1a* transcripts, whereas females lack the *Dmrt1* exon encoding the DM domain (Domingos et al. 2018). High-temperature feminisation in barramundi is likewise accompanied by changes in DNA methylation and alternative splicing of these genes (Budd et al. 2026). Comparative analyses across teleost fishes highlight a broad diversity of *Dmrt1* transactivation regions and isoforms with sex and stage-specific expression, suggesting rich potential for regulatory divergence at this crucial node (Dong et al. 2020; Akashi et al. 2024; Augstenová and Ma 2025). For amphibians, evidence is more fragmentary, but studies on *Dmrt1* have not shown any alternative splicing events (Augstenová and Ma 2025). Overall, ectothermic vertebrates provide several cases where environment-responsive alternative splicing acts near key regulatory nodes and may help direct gonadal fate, rather than simply fine-tuning an already established sexual developmental pathway.

Taken together, the comparative data suggest that alternative splicing is pervasive, but the placement of alternative splicing within the developmental GRN differs in ways that align with the broader developmental framework proposed in this review. In mammals and birds, alternative splicing primarily tunes non-essential regulators in a GRN that is already constrained by an early, transcriptionally defined somatic master switch (*Sry→Sox9* or *Dmrt1* dosage), by a pre-established dosage-compensation landscape and by tight temporal canalisation of the decision window (Zhao et al. 2018; Gómez-Redondo et al. 2021; Ioannidis et al. 2021). In many ectothermic vertebrates with labile sex determination, alternative splicing sits closer to the top master signals which are environmentally sensitive (e.g. *Rmb20–Wt1*, *Dmrt1*, *Cyp19a1a*) and can be directly modulated by temperature or social environment during a prolonged sensitive period (Domingos et al. 2018; Whiteley et al. 2022; Sun et al. 2025). Within this context, we propose that alternative splicing contributes to the developmental GRN lock-in framework. In lineages where sex fate is determined partly by alternative splicing decisions near top regulatory nodes, especially where those splicing events integrate environmental signals, the developmental GRN has a larger mutational and regulatory target for evolutionary changes, potentially facilitating rewiring of sex determination and sex-chromosome turnover. In lineages where alternative splicing acts mainly on non-essential nodes of an early, transcriptionally defined, dosage-compensated master switch, the GRN is developmentally more constrained, and repeated replacement of the master regulator is less probable.

## 7. How can the developmental GRN lock-in hypothesis be tested?

The developmental GRN lock-in hypothesis is proposed as a testable extension of existing models, not as a replacement for them. It can be evaluated alongside hypotheses based on sexually antagonistic selection, sex-ratio selection, meiotic drive, mutation load, the evolutionary-trap hypothesis, and chromosome-specific drift under stabilizing selection (Table 1) (Werren and Beukeboom 1998; Van Doorn and Kirkpatrick 2007; Van Doorn and Kirkpatrick 2007; Van Doorn and Kirkpatrick 2010; Blaser et al. 2013; Blaser et al. 2014; Ubeda et al. 2015; Muralidhar 2026). Its central prediction is that sex chromosome turnover should be least likely when the master sex-determining signal acts early, transiently, and in a specific somatic cell type, and when this signal is embedded in a canalized GRN coupled to chromosome-wide or otherwise strong dosage regulation. By contrast, turnover should be more likely when sexual fate remains plastic over a broader sensitive window, integrates environmental or physiological inputs, and is weakly coupled to the ancestral sex-chromosome dosage regulation.

This hypothesis can be tested by combining turnover histories with developmental and genomic variables across taxa. Relevant variables include sex chromosome age, homomorphy or heteromorphy, degree of Y/W degeneration, identity of the MSD, timing and cell type of master-regulator activity, duration of the sex-determining window, genomic scope of dosage compensation, environmental sensitivity or sex reversal, germ-cell involvement in primary gonadal fate decisions, and evidence for alternative splicing or epigenetic regulation near key GRN nodes. These variables could be analysed using phylogenetic generalized linear models, Bayesian threshold models, or phylogenetic mixed models to test whether developmental architecture improves prediction of turnover after accounting for heteromorphy, degeneration, mutation load, sexually antagonistic selection, meiotic drive, chromosome-specific drift and phylogenetic non-independence.

A critical test is whether developmental variables explain variation beyond existing models. For example, if heteromorphic sex chromosomes suppress turnover mainly when combined with early dosage regulation and an entrenched somatic master switch, this would support an interaction between evolutionary-trap dynamics and developmental GRN lock-in. Conversely, if turnover rates are fully explained by heteromorphy, degeneration, mutation load, or chromosome-specific drift, with no additional effect of developmental timing, cell type, or dosage-regulatory architecture, this would weaken the GRN lock-in hypothesis.

## 8. Conclusions and future directions

Sex chromosome turnover occurs rapidly in some vertebrate lineages yet remains rare or absent for >100 Myr in others. Existing models explain important parts of this contrast. The evolutionary trap hypothesis, and classical models centred on sexually antagonistic selection, mutation load, genetic drift, selfish genetic element accumulation (including meiotic drive), and more recent chromosome-specific drift under stabilizing selection all identify mechanisms that can either promote or constrain turnover. However, these models often treat the sex-determining locus largely as a genetic switch and give less attention to how the switch is embedded in development.

Here, we propose the developmental GRN lock-in hypothesis as a complementary framework. We argue that the probability of sex chromosome turnover depends not only on whether a new sex-determining locus can invade a population, but also on whether it can be developmentally accommodated. In therian mammals and birds, sex is specified by an early, transient, strictly genetic signal, such as *Sry*→*Sox9* or *Dmrt1* dosage, acting in somatic cells within a thermally buffered embryo. This signal is embedded in a canalised GRN and coupled to chromosome-scale or otherwise strong dosage regulation. In many ectothermic vertebrates, by contrast, sex-determining GRNs appear more flexible and integrative: genetic factors, germ cell number, epigenetic regulators, temperature, endocrine state, and other environmental cues may act over a broader sensitive window, with alternative splicing and local dosage regulation operating near key regulatory nodes.

We therefore suggest that in endothermic lineages with ancient and strongly differentiated sex chromosomes, the combination of an entrenched somatic master switch, early dosage regulation, and sex-linked regulatory integration creates a high barrier for invasion by alternative sex-determining loci. In many ectothermic lineages, broader developmental windows, environmental responsiveness, weaker or more local dosage regulation, and multiple potential regulatory entry points may make the sex-determining GRN more permissive to alternative upstream triggers. This does not imply that developmental GRN lock-in replaces existing models. Rather, it predicts that population-genetic forces such as sexually antagonistic selection, mutation, meiotic drive, or chromosome-specific drift will have different outcomes depending on the developmental architecture in which new sex-determining alleles arise.

At present, broad statistical tests of this hypothesis are limited by the lack of comparable developmental, genomic, and dosage-compensation data across vertebrates. We therefore propose a set of variables that future studies can score explicitly, including the timing and cell type of master-regulator action, width of the sex-determining window, role of germ cells, environmental sensitivity, sex chromosome differentiation, dosage-compensation scope, and the position of alternative splicing or epigenetic regulation within the GRN. These variables can be incorporated into phylogenetic comparative models to test whether developmental architecture explains turnover beyond existing predictors such as heteromorphy, degeneration, sexually antagonistic selection, mutation load, and meiotic drive. Testing the developmental GRN lock-in hypothesis will require isoform-resolved and cell-type–resolved transcriptomics across the sex-determining window, comparative developmental data from a broad range of taxa, and functional experiments targeting GRN nodes that are plausible candidates for evolutionary transitions. By integrating developmental GRNs with population genetics and comparative genomics, this framework aims to move the field towards quantitative, testable models that explain not only how sex chromosomes change, but also when and why some systems remain remarkably stable while others are repeatedly replaced.

### Author Contributions

W.-J. M. & Q. Z. designed the study. R. F. and T. C. collected data for sex chromosome turnover and alternative splicing respectively. W.-J. M. and P. V. synthesized ideas from the literature. P. V., R. F. and W.-J. M designed the figure visualization. W.-J. M. drafted the manuscript with significant input from P. V. and Q. Z. All authors have read and agreed to the submitted version of the manuscript.

### Data Availability Statement

NA.

### Acknowledgments

We gratefully acknowledge the Editor-in-Chief of GBE for inviting us to contribute this review, which grew out of a symposium we organized in SMBE 2025. We also thank three anonymous reviewers for their constructive feedback to improve the earlier version of the manuscript. This work is supported by the European Research Council (ERC) Starting Grant (FrogWY, 101039501), a Research Council grant (OZR4049) from Vrije Universiteit Brussel, and a research grant (G0A3B26N) from the Research Foundation Flanders (Fonds Wetenschappelijk Onderzoek, FWO), awarded to Wen-Juan Ma. The views and opinions expressed are those of the author(s) only and do not necessarily reflect those of the European Union or the European Research Council Executive Agency. Neither the European Union nor the granting authorities can be held responsible for them.

## Conflicts of Interest

The authors declare no conflict of interest.

## Data Availability Statement

No new data were generated or analysed in support of this manuscript.